\documentclass[aps,prx,groupedaddress,twocolumn,showpacs,longbibliography,10pt]{revtex4-2}
\DeclareFontShape{OT1}{cmr}{mx}{n}{<->cmr10}{}
\fontseries{mx}\selectfont
\usepackage{color}
\usepackage{bm} 
\usepackage{units}
\usepackage{bbm}
\usepackage[caption=false]{subfig}
\usepackage{graphicx}
\usepackage{dsfont}
\usepackage{amsmath}
\usepackage{amssymb}
\usepackage{array}
\usepackage{epstopdf}
\usepackage{enumerate}
\usepackage{youngtab}
\usepackage{tensor}
\usepackage{braket}
\usepackage[normalem]{ulem}
\usepackage{slashed}
\usepackage[aligntableaux=center]{ytableau}
\usepackage[utf8]{inputenc}
\usepackage[
      colorlinks=true,
      linkcolor=blue,
     urlcolor=blue,
      filecolor=black,
      citecolor=red,
      ]{hyperref}
\usepackage{cleveref}
\usepackage{braket}
\usepackage{mathtools}
\usepackage{rotating}
\usepackage{tabularx}
\newcolumntype{Y}{>{\centering\arraybackslash}X}
\newcolumntype{C}[1]{>{\centering\arraybackslash}p{#1}}
\usepackage{todonotes}
\usepackage{xcolor}
\definecolor{LightCyan}{rgb}{0.7,1,1}
\definecolor{Gray}{gray}{0.9}
\usepackage{xspace}

\begin{document}

\title{Magic Angle Butterfly in Twisted Trilayer Graphene}

\author{Fedor K. Popov and Grigory  Tarnopolsky}
\affiliation{Department of Physics, New York University, New York, NY 10003, USA}
\affiliation{Department of Physics, Carnegie Mellon University, Pittsburgh, PA 15213, USA}

\begin{abstract}
We consider a configuration of three stacked graphene monolayers 
with  commensurate twist angles $\theta_{12}/\theta_{23}=p/q$, where  $p$ and $q$ are coprime integers with $0<p<|q|$ and $q$ can be positive or negative. We study this system using the continuum model in the chiral limit when 
interlayer coupling terms between $\textrm{AA}_{12}$ and $\textrm{AA}_{23}$ sites of  the moir\'{e} patterns $12$ and $23$ are neglected. There are only three inequivalent displacements between the moir\'{e} patterns $12$ and $23$, at which the three monolayers' Dirac zero modes are protected. Remarkably, for these displacements and an arbitrary $p/q$ we discover exactly flat bands  at an infinite set of twist angles (magic angles). We provide theoretical explanation and classification of all possible configurations and topologies of the flat bands.

\end{abstract}


\maketitle
\nopagebreak

\section{Introduction and Summary}

Two graphene layers placed one on top of the other with a relative small twist angle form a periodic moir\'{e} pattern, which alters significantly the low energy electronic spectrum. At the special twist "magic" angle  $\theta_* \approx 1.05^\circ$
a dramatic flattening of the lowest energy bands was observed in \cite{doi:10.1073/pnas.1108174108, PhysRevB.82.121407}. These almost dispersionless electronic  bands hinted to a possibility of interesting  strongly interacting phenomena, which was subsequently confirmed through a series of experimental studies
\cite{CaoFatemiNature2018, CaoFatemiNature2, doi:10.1126/science.aav1910}.
These tantalizing experiments have inspired numerous theoretical and experimental investigations  \cite{PhysRevX.8.031089, PhysRevB.98.075109, PhysRevB.98.085435, doi:10.1073/pnas.1810947115, PhysRevB.98.085144, PhysRevB.98.195101, PhysRevLett.122.026801, PhysRevLett.122.086402, PhysRevB.98.035404, PhysRevB.98.045103, PhysRevLett.121.087001, PhysRevB.98.081102, PhysRevLett.121.257001, PhysRevB.99.075127,  PhysRevX.8.031088, Pizarro_2019, PhysRevX.8.031087, PhysRevB.98.241407, PhysRevX.8.041041, PhysRevB.98.235158, PhysRevB.98.220504,PhysRevB.99.144507, PhysRevB.106.235157, PhysRevLett.122.106405, PhysRevLett.108.216802, PhysRevLett.123.036401, PhysRevB.99.035111, PhysRevB.99.195455, PhysRevX.13.021012}, and the field continues to advance with fascinating new proposals and insights.

 Multiple layers of graphene stacked on top of each other with small relative twist angles \cite{PhysRevB.100.085109, PhysRevLett.123.026402, CeaWaletGuinea2019, PhysRevLett.125.116404, mao2023supermoire, lin2022energetic, ma2023doubled, PhysRevB.105.195422,LeeKhalafShangNature2019, PhysRevLett.128.176404, ledwith2021tb, PhysRevLett.128.176403, ZhangXieWu2023, yang2023flat} provide even greater versatility compared to Twisted Bilayer Graphene (TBG) due to the increased number of parameters. Initial theoretical investigations of alternating-twist trilayer graphene (aTTG)  \cite{PhysRevB.100.085109} demonstrated a similar flattening of electronic bands at "magic" angles, ultimately leading to the experimental discovery of various correlated phenomena \cite{PhysRevLett.127.166802, ParkCaoJarilloNature2021, doi:10.1126/science.abg0399,  LiuLiNaturePhys2022, doi:10.1126/science.abk1895, uri2023superconductivity} at the angles predicted in \cite{PhysRevB.100.085109}. 
 The study of the interaction effects in such systems remains an active area of ongoing theoretical investigations \cite{PhysRevB.103.195411, PhysRevB.104.115167, PhysRevX.12.021018}.

\begin{figure}[h!]
\centering
\includegraphics[scale=0.16]{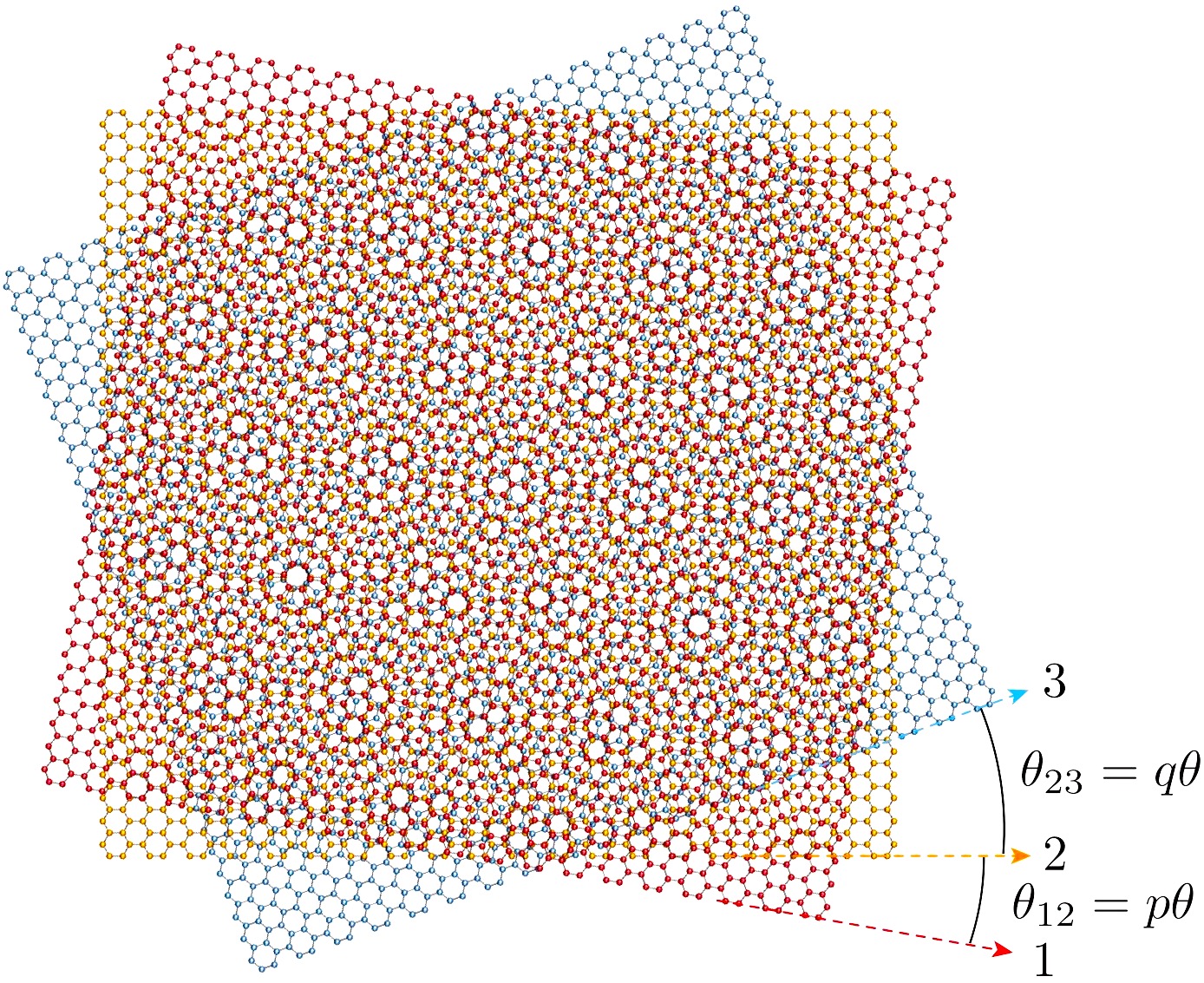}
\caption{Picture of a pq-Twist Trilayer Graphene (pqTTG) with $(p,q)=(1,2)$.  }
\label{fig:pqTTG}
\end{figure}

In \cite{PhysRevLett.123.026402, mao2023supermoire} a general configuration of twisted trilayer graphene 
was proposed, where the three layers are consecutively twisted by small angles $\theta_{12}=p\theta$ and $\theta_{23}=q\theta$, where $p$ and $q$ are coprime integers $0<p<q$, as shown in Fig. \ref{fig:pqTTG}. 
The analysis of electronic bands in this system,  was initially performed using the continuum model with equal interlayer coupling terms between $\textrm{AA}$ and $\textrm{AB}$ sites of the moiré patterns. However, this model did not exhibit a distinct phenomenon of band flattening.

The existence of magic angles and perfectly flat bands was recently discovered in the case of equal-twist trilayer graphene (eTTG) \cite{popov2023magic}, where the twist angles between layers 1 and 2 ($\theta_{12}$) and between layers 2 and 3 ($\theta_{23}$) are equal. The chiral limit of the twisted graphene continuum model  \cite{PhysRevLett.122.106405, PhysRevLett.108.216802} was crucial in revealing these magic angles and flat bands. In this limit, the interlayer coupling terms between $\textrm{AA}$ sites in the moiré pattern are disregarded, resulting in the Hamiltonian exhibiting exactly flat bands at an infinite series of magic angles.

In addition to the twist angles, the spectrum of a general twisted trilayer graphene (TTG) system is influenced by the relative displacement $\textbf{d}$ between the two moir\'{e} patterns formed by layers $12$ and $23$ \cite{mao2023supermoire}. The investigation of equal-twist trilayer graphene (eTTG) without any displacement between the moir\'{e} patterns ($\textbf{d}=0$, AAA stacking) revealed an intriguing connection between the flat bands of twisted bilayer graphene  and eTTG, establishing a relation between the magic angles in these distinct systems. A subsequent study by Guerci et al. \cite{guerci2023chern} (see also \cite{devakul2023magic, nakatsuji2023multi,foo2023extended}) demonstrated that  eTTG system exhibits a "moir\'{e} of moir\'{e}" pattern, resulting in large triangular regions of ABA and BAB stacking ($\textbf{d}= \pm \frac{1}{3}(\textbf{a}_{1}-\textbf{a}_{2})$, where $\textbf{a}_{1,2}$ are the single moir\'{e} lattice unit vectors) separated by smaller AAA regions. By employing the chiral model with ABA or BAB stacking  of the graphene layers, another set of magic angles and perfectly flat bands were unveiled. Furthermore, it was shown that the flat bands in
eTTG can possess the Chern number $C=2$, and the mathematical origin of such flat bands was explained.

In this letter, we analyze the general  twisted trilayer graphene configuration in which the consecutive layers are twisted at small but commensurate angles  $\theta_{12}=p\theta$ and $\theta_{23}=q\theta$, where $p$ and $q$ are coprime integer numbers $0<p<|q|$ and $q$ can be positive or negative.  
We refer to this system as $pq$-Twist Trilayer Graphene (pqTTG), and it is schematically depicted in Fig. \ref{fig:pqTTG}. A configuration of this type has been recently achieved in an experiment \cite{uri2023superconductivity} and analyzed in \cite{foo2023extended}. We analyze the  continuum model for this trilayer graphene system and take the chiral limit, where  interlayer coupling terms are  $w_{\textrm{AB}}\approx 110 \textrm{meV}$ and $w_{\textrm{AA}}=0$.

\begin{figure}[t!]
\centering
\includegraphics[scale=0.73]{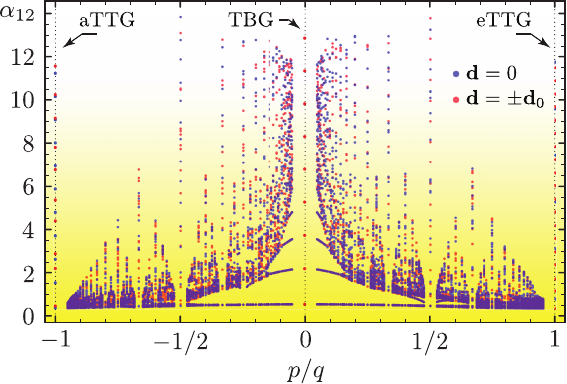}
\caption{The Magic Angle Butterfly. Plot of the leading magic angles $\alpha_{12}$ for all possible ratios  $p/q$ of coprime integers $p$ and $q$ with $0<p<|q|$ and up to $|q|=20$ and displacements $\textbf{d}=0, \pm \textbf{d}_{0}$. Near the center ($p/q=0$) the magic angles approach those of TBG.}
\label{fig:Butterfly}
\end{figure}

\begin{table}[t!]
\begin{center}
\begin{tabular}{l | c | c | c | c} 
\hline
$(p,q)_{\textbf{d}}$ & $\alpha_{12}^{(1)}$ & $\alpha_{12}^{(2)}$ & $\alpha_{12}^{(3)}$ & $\alpha_{12}^{(4)}$   \\ [0.5ex]
\hline
\hline
$(1,4)_{0}$ & $0.5673$ &   $1.4042$ & $ 1.8253 $   & $1.8295$ \\
\hline
$(1,4)_{\pm\textbf{d}_{0}}$ & $0.5672$ &   $1.6416_{4}$ & $ 1.7963 $   & $2.0260_{4}$ \\
\hline
$(1,3)_{0}$ & $0.5542$ &   $1.2025$ & $2.8487$ & $4.0074$  \\
\hline
$(1,3)_{\pm\textbf{d}_{0}}$ & $0.5527$ &   $1.2464$ & $1.3919$ & $2.0304$  \\
\hline
$(1,2)_{0}$ & $0.5194$ & $0.9589$ & $1.0255_{4}$   & $1.5309_{4}$   \\ 
\hline
$(1,2)_{\pm\textbf{d}_{0}}$ & $0.5094$ & $1.0698$ & $1.6226$   & $2.9886$   \\ 
\hline
$(2,3)_{0}$ & $0.4403$ &   $0.4549$ & $0.4954_{4}$  & $0.7662_{4}$ \\
\hline
$(2,3)_{\pm\textbf{d}_{0}}$ & $0.4527$ &   $0.4548$ & $0.4668$  & $0.5010$ \\
\hline
$(3,4)_{0}$ & $0.4356$ &   $0.4710$ & $0.4941_{4}$  & $0.5012$  \\
\hline
$(3,4)_{\pm\textbf{d}_{0}}$ & $0.4269$ &   $0.4302$ & $0.4612$  & $0.4646$  \\
\hline
$(1,1)_{0}$ & $0.8283_{4}$ & $3.1412_{4}$ & $5.3053_{4}$ & $7.4621_{4}$    \\ 
\hline
$(1,1)_{\pm\textbf{d}_{0}}$ & $0.3771$ & $1.1967_{4}$ & $1.7549$ & $2.4136_{4}$    \\ 
\hline
\end{tabular}
\end{center}
\caption{The table of magic angles for various $p,q>0$}
\label{fig:Table1}
\end{table}

\begin{figure}[t!]
\centering
\includegraphics[scale=0.83]{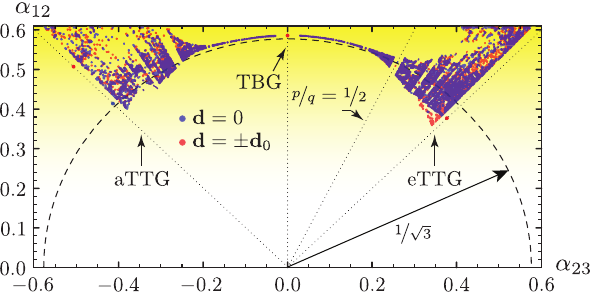}
\vspace{0.61cm}
\caption{Plot of the leading magic angles $(\alpha_{12},\alpha_{23})$  for  all possible ratios  $p/q$ of coprime integers $p$ and $q$ with $0<p<|q|$ and up to $|q|=20$. The black dashed line is a semicirlce $\alpha_{12}^{2}+\alpha_{23}^{2}=1/3$ derived in \cite{PhysRevLett.125.116404} using perturbation theory.}
\label{fig:Semicircle}
\end{figure}

For small twist angles we assume that the moir\'{e} patterns $12$ and $23$ are formed by vectors $p\textbf{a}_{1,2}$ and $q\textbf{a}_{1,2}$,  where $\mathbf{a}_{1,2} = (4\pi/3 k_\theta) (\pm \sqrt{3}/2, 1/2)$ with $k_{\theta}= 2k_{D}\sin(\theta/2)\approx k_{D}\theta$ and $k_{D}=4\pi/3\sqrt{3}a$ is the Dirac momentum of the monolayer graphene with lattice constant $a \approx 1.42$\AA. Similarly to the previously discussed eTTG configuration, pqTTG also has a moiré of moiré pattern, resulting in local variations of the displacement vector $\textbf{d}$ between the moir\'{e} patterns $12$ and $23$. There are only three inequivalent displacement vectors, at which  the three monolayers' Dirac zero modes are protected. These are  $\textbf{d}=0$ and $\textbf{d}=\pm \textbf{d}_{0}$, where $\textbf{d}_{0} = \frac{1}{3p|q|}(\textbf{a}_{1}-\textbf{a}_{2})$ and  the spectrum is invariant under the shifts $\textbf{d} \to \textbf{d} + \frac{1}{pq}\textbf{a}_{1,2}$. 
The values of the magic angles are identical for the displacements $\pm \textbf{d}_0$.  Remarkably, in the chiral limit and these displacements the electronic energy spectrum exhibits perfectly flat bands at an infinite sequence of magic angles for any combination of coprime integers $p$ and $q$.  We introduce dimensionless twist parameters $\alpha_{12}=\alpha/p$ and $\alpha_{23}=\alpha/q$ where $\alpha = w_{\textrm{AB}}/(v_{F}k_{D}\theta)$, and $v_{F}\approx 10^{6}$m/s is the  monolayer graphene Fermi velocity (for brevity we also refer to $\alpha_{12}$, $\alpha_{23}$ and $\alpha$ as twist angles). We plot magic angles $\alpha_{12}$ as a function of $p/q$  in Fig. \ref{fig:Butterfly}.  Plot of the magic angles $(\alpha_{12}$, $\alpha_{23})$ for different $p/q$ is shown in Fig. \ref{fig:Semicircle}.
Finally we present Tables \ref{fig:Table1} and \ref{fig:Table2} of the first four magic angles $\alpha_{12}$ for various values of $p$ and $q$ and displacements $\textbf{d}=0$ and $\textbf{d}=\pm \textbf{d}_{0}$.
\begin{table}[t!]
\vspace{0.1cm}
\begin{center}
\begin{tabular}{l | c | c | c | c} 
\hline
$(p,q)_{\textbf{d}}$ & $\alpha_{12}^{(1)}$ & $\alpha_{12}^{(2)}$ & $\alpha_{12}^{(3)}$ & $\alpha_{12}^{(4)}$   \\ [0.5ex]
\hline
\hline
$(1,-1)_{0}$ & $0.4141$ & $1.5706$ & $2.6526$   & $3.7310$   \\ 
\hline
$(1,-1)_{\pm\textbf{d}_{0}}$ & $0.5072$ & $2.2295$ & $2.9043$   & $4.4080$   \\ 
\hline
$(3,-4)_{0}$ & $0.4821$ & $0.4887_{4}$ & $0.6493$   & $0.6626_{4}$   \\ 
\hline
$(3,-4)_{\pm\textbf{d}_{0}}$ & $0.5121$ & $0.6167$ & $0.6520$   & $0.6954$   \\ 
\hline
$(2,-3)_{0}$ & $0.4912_{4}$ & $0.5054$ & $1.2890$   & $1.3259_{4}$   \\ 
\hline
$(2,-3)_{\pm\textbf{d}_{0}}$ & $0.9458$ & $1.0945$ & $1.7939$   & $2.2357$   \\ 
\hline
$(1,-2)_{0}$ & $0.5131$ & $1.5014$ & $2.0490$   & $3.2157$   \\ 
\hline
$(1,-2)_{\pm\textbf{d}_{0}}$ & $0.5857$ & $0.7614$ & $1.3797$   & $1.8178_{4}$   \\ 
\hline
$(1,-3)_{0}$ & $0.5574$ & $1.4373$ & $2.6738_{4}$   & $3.9906_{4}$   \\ 
\hline
$(1,-3)_{\pm\textbf{d}_{0}}$ & $0.5587$ & $1.2915$ & $1.3920$   & $2.1730$   \\ 
\hline
$(1,-4)_{0}$ & $0.5685$ & $1.7225_{4}$ & $1.8548_{4}$   & $3.2869$   \\ 
\hline
$(1,-4)_{\pm\textbf{d}_{0}}$ & $0.5685$ & $1.7130$ & $1.7298$   & $1.9467$   \\ 
\hline
\end{tabular}
\end{center}

\caption{The table of magic angles for various $p,-q>0$}
\label{fig:Table2}
\end{table}

Below we formulate the continuum model for twisted trilayer graphene,  present the criteria for the emergence of perfectly flat bands and provide a complete classification of the  structures of the flat bands at magic angles. 
\section{Continuum model for Twisted Trilayer Graphene}
\label{sec:contmodel}

We consider a system of three stacked graphene monolayers, where each layer $\ell=1,2,3$ is rotated counterclockwise by an angle $\theta_{\ell}$ around  an atom site and then shifted by a vector $\textbf{d}_{\ell}$, so atoms in
each layer are parametrized by $\textbf{r}=R_{\theta_{\ell}}(\textbf{R}+\tau_{\alpha})+\textbf{d}_{\ell}$, where $R_{\theta}= e^{-i\theta \sigma_{y}}$ is the 
rotation matrix and $\textbf{R}$ and $\tau_{\alpha}$ run over the lattice and sub-lattice sites. The continuum model Hamiltonian for twisted trilayer graphene can be written as \cite{PhysRevB.100.085109}:
\begin{align*}
H= \left( \begin{array}{ccc}
    -iv_{F}  \bm{\sigma}_{\theta_{1}}\bm{\nabla} & T^{12}(\textbf{r}-\textbf{d}_{12}) &0 \\ 
    T^{12\dag}(\textbf{r}-\textbf{d}_{12}) & -iv_{F}  \bm{\sigma}_{\theta_{2}}\bm{\nabla} & T^{23}(\textbf{r}-\textbf{d}_{23}) \\ 
    0 & T^{23\dag}(\textbf{r}-\textbf{d}_{23}) & -iv_{F}  \bm{\sigma}_{\theta_{3}}\bm{\nabla} \\ 
  \end{array} \right)\,, \label{HamFullInitial}
\end{align*}
where $\bm{\sigma}_{\theta}\equiv e^{i\frac{\theta}{2}\sigma_{z}} \bm{\sigma}e^{-i\frac{\theta}{2}\sigma_{z}}$, $ \bm{\sigma} = (\sigma_{x}, \sigma_{y})$ and $\textbf{d}_{\ell \ell'}=\frac{1}{2}(\textbf{d}_{\ell}+\textbf{d}_{\ell'} +i\cot(\theta_{\ell'\ell}/2)\sigma_{y}(\textbf{d}_{\ell}-\textbf{d}_{\ell'}))$ is the moir\'{e} pattern displacement vector. The moir\'{e} potential between adjacent layers $\ell$ and $\ell'$ is
\begin{align}
T^{\ell \ell'}(\textbf{r}) = \sum_{n=1}^{3}T^{\ell \ell'}_{n}e^{-i\textbf{q}_{n}^{\ell \ell'}\textbf{r}}\,, 
\end{align}
where $T^{\ell \ell'}_{n+1}=w_{\textrm{AA}}^{\ell \ell'}\sigma_{0}+w_{\textrm{AB}}^{\ell \ell'}(\sigma_{x} \cos n\phi +\sigma_{y}\sin n\phi)$ and 
\begin{align}
\textbf{q}_{1}^{\ell \ell'} = 2k_{D}\sin(\theta_{\ell'\ell}/2)R_{\phi_{\ell \ell'}}(0,-1), \;\; \textbf{q}^{\ell \ell'}_{2,3}=R_{\pm \phi}\textbf{q}^{\ell \ell'}_{1}
\end{align}
with $\theta_{\ell \ell'} =\theta_{\ell}-\theta_{\ell'}$,  $\phi_{\ell \ell'}=(\theta_{\ell}+\theta_{\ell'})/2$, $\phi=2\pi/3$. The coupling between adjacent layers $\ell$ and $\ell'$ is characterized by two parameters $w_{\textrm{AA}}^{\ell \ell'}$ and $w_{\textrm{AB}}^{\ell \ell'}$ representing intra- and intersub-lattice couplings. The chiral limit corresponds to 
 $w_{\textrm{AA}}^{\ell \ell'}=0$. Also using translation invariance we can make a replacement $\textbf{r}\to \textbf{r}+ \mathbf{d}_{12}$, and therefore the Hamiltonian depends only on a single displacement vector $\mathbf{d} = \mathbf{d}_{23} - \mathbf{d}_{12}$.

We consider a trilayer configuration where the relative twists are commensurate  
$\theta_{21}=p\theta$ and $\theta_{32}=q\theta$ and $p$ and $q$ are two coprime integers, which satisfy  $0<p<|q|$.
For a small angle $\theta$ we can set $\phi_{ll'} = 0$ leading to $\textbf{q}_1^{12} = p\textbf{q}_{1} $ and $\textbf{q}_1^{23} = q \textbf{q}_{1}$ and $\textbf{q}_{1}=k_{\theta}(0,-1)$ with $k_{\theta}= 2k_{D}\sin(\theta/2)\approx k_{D}\theta$. Thus we obtain the following Hamiltonian
\begin{align}
H_{\textrm{pqTTG}}= \left( \begin{array}{ccc}
    -iv_{F}  \bm{\sigma}_{-p\theta}\bm{\nabla} & T(p\textbf{r}) &0 \\ 
    T^{\dag}(p\textbf{r}) & -iv_{F}  \bm{\sigma}\bm{\nabla} & T(q(\textbf{r}-\textbf{d})) \\ 
    0 & T^{\dag}(q(\textbf{r}-\textbf{d})) & -iv_{F}  \bm{\sigma}_{q\theta}\bm{\nabla} \\ 
  \end{array} \right)\,, \label{eq:Hamtril}
 \end{align}
where we assume equal couplings between layers. For a small twist angle $\theta$ we can neglect the phase factors in the Pauli matrices $\bm{\sigma}_{-p\theta} \to \bm{\sigma}$ and $\bm{\sigma}_{q\theta} \to \bm{\sigma}$.

\begin{figure}
\centering
\includegraphics[scale=0.65]{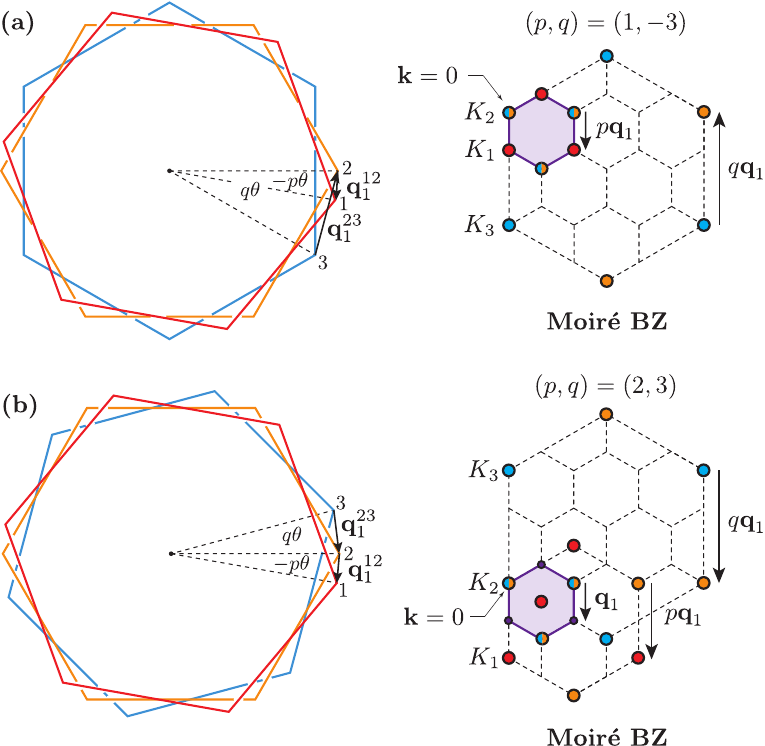}
\caption{Original Brillouin zones of three  graphene layers with their Dirac points $K_{1}$, $K_{2}$ and $K_{3}$ and the moir\'{e} Brillouin zones for  pqTTG. The layers are  twisted by the angles $p\theta$ and $q\theta$, where $p$ and $q$ are coprime integers such that $0<p<|q|$. The wave vector $\textbf{k}$ is zero at the Dirac point $K_{2}$. We neglect a relative rotation between vectors $\textbf{q}_{1}^{12}$ and $\textbf{q}_{1}^{23}$ and assume that $\textbf{q}_{1}^{12}=p\textbf{q}_{1}$ and $\textbf{q}_{1}^{23}=q\textbf{q}_{1}$. (a)  Case $(p,q)=(1,-3)$.   (b) Case $(p,q)=(2,3)$. }
\label{fig:TLGstruct}
\end{figure}

The moir\'{e} Brillouin zone (mBZ) for this Hamiltonian is depicted in Fig. \ref{fig:TLGstruct}. The  reciprocal moir\'{e} unit vectors are $\mathbf{b}_{1,2} = \mathbf{q}_{2,3}-\mathbf{q}_1$. In the coordinate space the unit vectors are $\mathbf{a}_{1,2} = (4\pi/3 k_\theta) (\pm \sqrt{3}/2, 1/2)$.  It is useful to introduce complex coordinates  $z, \bar{z}=\mathbf{r}_x \pm  i \mathbf{r}_y$ in real space and $k, \bar{k} = \mathbf{k}_1 \pm  i \mathbf{k}_2$ in momentum space.

The Hamiltonian (\ref{eq:Hamtril}) acts on a spinor $\Psi(\textbf{r})=(\psi_{1},\chi_{1}, \psi_{2},\chi_{2},\psi_{3},\chi_{3})$, where the indices $1,2,3$ represent the graphene layer. Introducing the dimensionless twist parameter $\alpha = w_{\textrm{AB}}/(v_{F} k_{\theta})$, and
writing the Hamiltonian \eqref{eq:Hamtril} in the sublattice basis $\Psi(\textbf{r})=(\psi_{1}, \psi_{2}, \psi_{3}, \chi_{1},\chi_{2},\chi_{3})$ we obtain
\begin{align}
\mathcal{H}_{\textrm{pqTTG}} = \begin{pmatrix}
 \mathcal{M}(\textbf{r}) & \mathcal{D}^{\dag}(\textbf{r})\\
\mathcal{D}(\textbf{r}) &  \mathcal{M}(\textbf{r})
\end{pmatrix}\,,
 \label{eq:pqTTG}
\end{align}
where we have rescaled coordinates $\textbf{r} \to k_{\theta}\textbf{r}$ and the Hamiltonian, so the energies of (\ref{eq:pqTTG}) are measured in units of $v_{F}k_{\theta}$. 
The operators $\mathcal{D}$ and $\mathcal{M}$  are 
\begin{align}
&\mathcal{D}(\textbf{r}) = \left(  \begin{array}{ccc}
    -2i\bar{\partial} & \alpha U_1(p \textbf{r}) & 0 \\ 
    \alpha U_1(-p\textbf{r}) & -2i\bar{\partial} & \alpha U_1(q(\textbf{r}-\textbf{d})) \\ 
    0 & \alpha U_1(q(\textbf{d}-\textbf{r})) & -2i\bar{\partial} \\  
  \end{array}\right)\,, \notag\\
&\mathcal{M}(\mathbf{r}) = \frac{w_{\textrm{AA}}}{w_{\textrm{AB}}} \left(  \begin{array}{ccc}
    0 &  U_0(p\textbf{r}) & 0 \\ 
     U_0(-p\textbf{r}) & 0 &  U_{0}(q(\textbf{r}-\textbf{d})) \\ 
    0 &  U_{0}(q(\textbf{d}-\textbf{r})) & 0 \\  
  \end{array}\right)\,,  \label{eq:D3andM}
\end{align}
where $\partial, \bar{\partial} = \frac{1}{2}(\partial_{x} \mp i \partial_{y})$ are complex derivatives and we defined $U_m(\mathbf{r}) = \sum^{3}_{n=1} \omega^{m(n-1)} e^{-i \textbf{q}_n \textbf{r}}$ with $\omega=e^{i\phi}$.  The Bloch states $\Psi_{\textbf{k}}(\textbf{r})=(\psi_{\textbf{k}}(\textbf{r}), \chi_{\textbf{k}}(\textbf{r}))$ of  \eqref{eq:pqTTG}  are parametrized by the wave vector $\textbf{k}$ from mBZ and satisfy the following boundary conditions
\begin{gather}
\Psi_\mathbf{k}(\mathbf{r}+\mathbf{a}_{1,2}) = e^{i \mathbf{k}\, \mathbf{a}_{1,2}}  U_\phi \Psi_\mathbf{k}(\mathbf{r})\,, \label{BCTTG}
\end{gather}
where the matrix $U_{\phi} = \mathds{1}_{\textrm{AB}}\otimes \operatorname{diag}\left(\omega^{p},1,\omega^{-q}\right)$.  

Below we focus on the chiral limit $w_{\textrm{AA}}=0$. In this case the Hamiltonian (\ref{eq:pqTTG}) is particle-hole symmetric $\{\mathcal{H}_{\textrm{pqTTG}}, \sigma_{z}\otimes  \mathds{1}\} =0$.  The chiral and anti-chiral zero energy equations are 
\begin{align}
\mathcal{D}(\textbf{r}) \psi_{\textbf{k}}(\textbf{r}) = 0, \quad \mathcal{D}^{\dag}(\textbf{r}) \chi_{\textbf{k}}(\textbf{r}) = 0\,, 
\end{align}
and they have solutions for every wave vector $\textbf{k}$ of the mBZ (flat bands)
only at the special values (magic angles) of the twist angle $\alpha$.  We can represent the chiral operator $\mathcal{D}(\mathbf{r})$ in (\ref{eq:D3andM}) in the form
\begin{gather}
\mathcal{D}(\mathbf{r})   = -2 i \bar{\partial} + \alpha \bar{A}(\textbf{r})\,,\label{opDexp}
\end{gather}
where we introduced the  matrix vector-potential $\bar{A}(\textbf{r})$. The magic angles $\alpha$ can be found numerically as a generalized eigenvalues of the operators $-2 i \bar{\partial}$ and $ \bar{A}(\textbf{r})$  \cite{becker2021spectral, becker2022mathematics}.  Generically the magic angles $\alpha$ form an infinite set of isolated points in the entire complex plane. In order to have magic angles, it is necessary for the chiral operator $\mathcal{D}(\textbf{r})$ to adhere to specific symmetries.
Namely we would like the chiral operator to have three Dirac zero modes $\psi_{K_1}$, $\psi_{K_2}$ and $\psi_{K_3}$ at the Dirac points $K_{1}=p\textbf{q}_{1}$,  $K_{2}=0$ and $K_{3}=-q \textbf{q}_{1}$ for an arbitrary twist angle $\alpha$. This is guaranteed 
by $C_{3z}$ symmetry together with the particle-hole symmetry \cite{PhysRevLett.122.106405, PhysRevLett.123.026402}. 
A non-zero displacement $\textbf{d}$ can break the $C_{3z}$ symmetry of the Hamiltonian.  One can show that  spectrum of the Hamiltonian  (\ref{eq:pqTTG}) is invariant under the  
following shifts of the displacement 
\begin{align}
    \mathbf{d} \to \mathbf{d}  + \frac{1}{pq}\textbf{a}_{1,2}\,
\end{align}
and the inequivalent displacements that preserve the three Dirac zero modes are $\textbf{d} =0$ and  $\textbf{d} = \pm \textbf{d}_0$, where 
\begin{gather}
  \textbf{d}_{0} =  \frac{1}{3p|q|}\left(\textbf{a}_1-\textbf{a}_2\right)\,.
\end{gather}
The magic angles of the the Hamiltonian are identical for the displacements $\textbf{d}=\pm\textbf{d}_0$, because they are related to each other by the $C_{2z}\mathcal{T}$ symmetry, 
that exchanges the chiral and anti-chiral components of the Bloch states.

\section{Origin of the flat bands}
In this section we  derive  general properties of the zero-modes of the  Hamiltonian \eqref{eq:pqTTG} in the chiral limit $w_{\textrm{AA}}=0$.
As we discussed above, the chiral operator admits three Dirac zero modes $\psi_{K_1}$, $\psi_{K_2}$ and $\psi_{K_3}$ at the Dirac points $K_{1}=p\textbf{q}_{1}$,  $K_{2}=0$ and $K_{3}=-q \textbf{q}_{1}$  
\begin{gather}
\mathcal{D}(\mathbf{r})\psi_{K_{i}}(\mathbf{r}) = 0, \quad i = 1,2,3 \label{DirZM}
\end{gather}
for the displacements $\mathbf{d}=0, \pm \mathbf{d}_{0}$ and an arbitrary twist angle $\alpha$.
Using that $\textrm{Tr}\bar{A}(\textbf{r})=0$ and equations (\ref{DirZM}) one can derive that the Wronskian of the Dirac spinors  
\begin{gather}
W(\mathbf{r}) \equiv \det \left(\psi_{K_{1}}, \psi_{K_{2}}, \psi_{K_3} \right)  \label{Wr1}
\end{gather}
satisfies  $\bar{\partial} W(\mathbf{r}) = 0$ \cite{PhysRevB.103.155150},\cite{guerci2023chern}. (Note that the Wronskian in this case is simply a triple product: $W(\textbf{r})= \psi_{K_{1}} \cdot (\psi_{K_{2}}\times \psi_{K_{3}})$).
Therefore we must conclude that the Wronskian is a constant $W(\textbf{r})=W$, since we can not have a non-constant holomorphic function on a compact manifold, which is a torus in our case.  

First we prove that  if $W \neq 0$ there is no flat band. $W \neq 0$ implies that $\psi_{K_i}(\textbf{r})$, $i=1,2,3$ as $3$-dimensional vectors are linearly independent at each point $\mathbf{r}$ of the moir\'{e} unit cell, and form a basis. Therefore any other chiral zero mode $\psi_\mathbf{k}(\mathbf{r})$  ($\mathcal{D}\psi_{\textbf{k}} = 0$) can be expanded as 
\begin{gather}
\psi_\mathbf{k}(\mathbf{r}) = \mathcal{C}_1(\mathbf{r}) \psi_{K_{1}}(\mathbf{r})+
 \mathcal{C}_2(\mathbf{r}) \psi_{K_{2}}(\mathbf{r})+ \mathcal{C}_3(\mathbf{r}) \psi_{K_{3}}(\mathbf{r})\,, \notag
\end{gather}
where the scalar functions $ \mathcal{C}_{1,2,3}$ must depend on the wave vector $\textbf{k}$ in order for $\psi_\mathbf{k}$ to satisfy the the Bloch boundary conditions (\ref{BCTTG}). 
Applying  the operator $\mathcal{D}(\mathbf{r})$ to both parts of the above equality we conclude that $\bar{\partial }\mathcal{C}_{1,2,3}(\mathbf{r})=0$.  And since $\psi_\mathbf{k}$ and $\psi_{K_{i}}$ are finite everywhere $\mathcal{C}_i(\mathbf{r})$ can only be constants. But then  $\psi_{\textbf{k}}$ would violate the boundary conditions (\ref{BCTTG}). Therefore  if $W\neq 0$ we can not have other zero energy solutions (and thus flat band) except the Dirac zero modes.

Now, we prove that if $W = 0$, we necessarily have a flat band. $W =0 $ implies that the Dirac spinors $\psi_{K_i}(\mathbf{r})$ as vectors are linearly dependent at every point $\mathbf{r}$ of the moir\'{e} unit cell. We stress that it does not mean that they are linearly dependent functions in the Hilbert space. There are two possible scenarios. 

\textit{Scenario 1.} All  Bloch spinors of the  zero energy flat bands are collinear at every point $\mathbf{r}$ of the moir\'{e} unit cell. We  refer to such case as rank 1 flat bands. In this situation any flat band's wave spinor can be described by the following equation:
\begin{gather}
    \psi_{\mathbf{k}}(\mathbf{r}) = \mathcal{C}(\mathbf{r}) \psi_{K_1}(\mathbf{r})\,, \label{eqRank1}
\end{gather}
where $\mathcal{C}(\mathbf{r})$ is some scalar function. 
If we apply the operator $\mathcal{D}(\textbf{r})$ to both sides of this equation we obtain $\bar{\partial } \mathcal{C}(\mathbf{r})=0$ and thus $ \mathcal{C}$ is a constant, which  violates the boundary conditions  (\ref{BCTTG}) for the spinor $\psi_{\textbf{k}}$. 
Therefore the eq. (\ref{eqRank1})  is impossible unless $\psi_{K_1}(\mathbf{r})$ has zeros. If $\psi_{K_1}(\mathbf{r})$ has $n$ zeros at the points $\textbf{r}_\lambda$, $\lambda=1,\dots,n$, we can construct $n$ flat bands, using the following wave functions: 
\begin{gather}
    \psi^{(\lambda)}_\mathbf{k}(\mathbf{r}) =  f_{\textbf{k} - K_1}(z - z_\lambda)  \psi_{K_1}(\mathbf{r})\,, \label{Rank1FB}
\end{gather}
where $z_\lambda = \left(\mathbf{r}_\lambda\right)_x + i \left(\mathbf{r}_\lambda\right)_y$ and we introduced the following meromorphic function
\begin{align}
 f_{\textbf{k}}(z) = e^{i \frac{\textbf{k}\textbf{a}_{1}}{a_{1}}z } \frac{\vartheta_{1}(z/a_1-k/b_2|\omega) }{\vartheta_{1}(-k/b_2|\omega) \vartheta_{1}(z/a_{1}|\omega)}\,. \label{eq:Ffunction}
\end{align}
Normalization of the function $f_{\textbf{k}}(z)$ is chosen such that $f_{\textbf{k}+\textbf{b}_{1,2}}(z)= f_{\textbf{k}}(z)$ and one can compute that the Chern number of the flat bands in (\ref{Rank1FB}) is $C= 1$.

\textit{Scenario 2.} The other case is that all wave spinors of the flat bands form a two-dimensional vector space at every point $\mathbf{r}$ of the moir\'{e} unit cell. We  refer to such a case as rank 2 flat bands. In this case we have 
\begin{gather}
\psi_{K_3}(\mathbf{r}) =  \mathcal{C}_{1}(\mathbf{r}) \psi_{K_1}(\mathbf{r})+ 
 \mathcal{C}_{2}(\mathbf{r}) \psi_{K_2}(\mathbf{r})\,,\label{eq:flatrel}
\end{gather}
where $\mathcal{C}_{1}(\mathbf{r})$ and $\mathcal{C}_{2}(\mathbf{r})$ are some scalar functions. 
If we assume that the spinors $\psi_{K_1}(\textbf{r})$ and $\psi_{K_2}(\textbf{r})$   are linearly independent at every point $\textbf{r}$ of the moir\'{e} unit cell and apply the operator $\mathcal{D}(\textbf{r})$ to both sides of the equation (\ref{eq:flatrel}) we obtain $\bar{\partial } \mathcal{C}_{1,2}(\mathbf{r})=0$ and thus $ \mathcal{C}_{1,2}$ are constants, which  violates the boundary conditions  (\ref{BCTTG}) for the spinor $\psi_{K_{3}}$. Therefore we conclude that there exists at least one point $\textbf{r}_{0}$ where the vectors $\psi_{K_1}(\mathbf{r}_{0})$ and $\psi_{K_2}(\mathbf{r}_{0})$  are linearly dependent. More generally, there could be
$n$ such points $\mathbf{r}_{\lambda}$, $\lambda=1,\dots, n$, where 
\begin{gather}
 c_{1\lambda}\psi_{K_{1}}(\mathbf{r}_\lambda)+ c_{2\lambda}\psi_{K_{2}}(\mathbf{r}_\lambda)= 0\,,   \label{eq:linpoint}
\end{gather}
for $ \lambda=1,\dots,n$.  (Notice that the coefficients $c_{1,2\lambda}$ are not equal to $\mathcal{C}_{1,2}(\textbf{r}_{\lambda})$). 
Now it is easy to see that  for each zero point $\mathbf{r}_\lambda$ we can construct a chiral zero mode at every point $\textbf{k}$ of the mBZ:
\begin{align}
\psi^{(\lambda)}_\mathbf{k}(\mathbf{r}) =  
\big( c_{1\lambda} &f_{\mathbf{k} - K_1}(z - z_\lambda) \psi_{K_{1}}(\mathbf{r})\label{eq:FunCon} \\
&\qquad\quad  +c_{2\lambda} f_{\mathbf{k} - K_2}(z - z_\lambda) \psi_{K_{2}}(\mathbf{r})\big)\,.\notag
\end{align}
The function $f_{\textbf{k}}$ defined in (\ref{eq:Ffunction}) behaves as 
\begin{align}
    f_{\textbf{k}}(z-z_\lambda) \to  \frac{a_1}{\vartheta_{1}'(0|\omega)} \frac{1}{z-z_\lambda}, \quad  z \to z_\lambda
\end{align}
and  the poles of the functions $f_{\mathbf{k}-K_{1}}(z)$ and $f_{\mathbf{k}-K_{2}}(z)$ are cancelled at the point $\mathbf{r}_{\lambda}$, thus $\psi_{\textbf{k}}^{(\lambda)}$ is finite  at every point of the moir\'{e} unit cell.  A similar construction within the context of eTTG was  proposed by Guerci et al. \cite{guerci2023chern}.

Since the function $f_{\textbf{k}}$ in \eqref{eq:Ffunction} is periodic in the mBZ, but  has a pole at $\mathbf{k}=0$  the function $\psi^{(\lambda)}_\mathbf{k}$ can have two poles in mBZ if both coefficients $c_{1\lambda}$ and $c_{1\lambda}$  are not zero.  In this case the flat band has the Chern number  $C=2$. 
If either of the coefficients $c_{1\lambda}$ or $c_{2\lambda}$ is zero, the function $\psi^{(\lambda)}_\mathbf{k}$ has one simple pole within the mBZ, we have a rank $1$ flat sub-band with the Chern number  $C=1$.

Thus, the rank 2 flat band is characterized by two wave functions $\psi_{K_{1}}(\mathbf{r})$ and $\psi_{K_{2}}(\mathbf{r})$. These wave spinors are linearly independent vectors   at every point in the moir\'{e} unit cell besides $n$ isolated points $\textbf{r}_{\lambda}$. Among these points, there may exist $n_1$ points where either of the spinors $\psi_{K_{1}}$ or $\psi_{K_{2}}$ is zero. These  points give rise to $n_1$ flat bands with the Chern number  $C=1$. At the rest of $n_2=n-n_1$ points  $\psi_{K_{1}}$ and $\psi_{K_{2}}$ are non-zero and linearly dependent. These points give rise to $n_2$  flat bands with the Chern number $C=2$.

We note that the above construction can be easliy generalized for the case of  twisted $\ell$-layer graphene in the chiral limit and shows that one can have flat bands with the Chern number up to $C=\ell-1$.

Finally we note that we can have only a single Dirac cone on top of  the flat bands. To demonstrate this, we observe that if there are two Dirac cones on top of the flat bands, their zero energy wave spinors must be linearly dependent at some point. Consequently, these wave functions would generate a rank 2 flat band, contradicting the assumption that they constitute distinct zero-energy states. 

Below we elaborate more on the possible structures of the flat bands in pqTTG. For this we separately consider  cases of $\textbf{d}=0$ and  $\textbf{d}=\pm\mathbf{d}_0$ displacements. We summarize all possible configurations of the flat bands  in the Fig. \ref{fig:strfld0} and  \ref{fig:strfld1}.

\section{Structures of the flat bands}
 In this section we discuss configurations of the flat bands and their Chern numbers for two different cases of the displacement: $\textbf{d}=0$ and $\textbf{d}=\pm \mathbf{d}_0$.

\subsection{Displacement $\textbf{d}=0$}
\textit{Absence of the $C=2$ flat bands.} In the case of zero displacement one can construct anti-chiral zero modes $\chi_{\textbf{k}}$ ($\mathcal{D}^{\dag}\chi_{\textbf{k}}=0$) from   chiral zero modes  $\psi_{\mathbf{k}}$ ($\mathcal{D}\psi_{\textbf{k}}=0$) as 
\begin{align}
    &\chi_{\mathbf{k}_{1,2}}(\mathbf{r}) = Q \bar{\psi}_{\mathbf{k}_{1,2}}(-\mathbf{r})\,\notag \\  &\chi_{\mathbf{k}}(\mathbf{r}) = Q\left(  \bar{\psi}_{\mathbf{k}_{1}}(\mathbf{r})\times \bar{\psi}_{\mathbf{k}_{2}}(\mathbf{r})\right), \label{eq:maps}
\end{align}
where $Q=\textrm{diag}(1,-1,1)$, $\mathbf{k} = \mathbf{k}_{*} - \mathbf{k}_1 - \mathbf{k}_2$ and $\mathbf{k}_{*} = (p-q)\mathbf{q}_1$.  By combining the above formulas
we can also construct an additional chiral zero mode $\psi_{\mathbf{k}}$ at the point $\mathbf{k} = \mathbf{k}_{*} - \mathbf{k}_1 - \mathbf{k}_2$: 
\begin{gather}
    \psi_{\mathbf{k}}(\mathbf{r}) =\psi_{\mathbf{k}_1}(-\mathbf{r}) \times \psi_{\mathbf{k}_2}(-\mathbf{r})\,. \label{eq:prodformula}
\end{gather}
Moreover one can check that that the following function
\begin{gather}
    v(\mathbf{r})=   \psi_{\mathbf{k}_1}(\mathbf{r})\cdot \psi_{\mathbf{k}_2}(-\mathbf{r}) \label{vhol}
\end{gather}
satisfies $\bar{\partial} v(\mathbf{r}) = 0$ \cite{PhysRevLett.122.106405}, therefore to obey the Bloch boundary conditions $v$ must be zero if $\mathbf{k}_{1} \neq  \mathbf{k}_{2}$.

Now let us assume that we have a rank $2$ flat band discussed in the \textit{Scenario 2} in the previous section. This means  that we have two wave spinors $\psi_{\mathbf{k}_1}$ and $\psi_{\mathbf{k}_2}$ that are linearly independent as vectors at every point in the moir\'{e} unit cell besides $n$ isolated points. By virtue of \eqref{eq:prodformula} we have an additional solution $\psi_{\mathbf{k}}$. Let us take a point $\textbf{k}_{3}$ in the mBZ such that $\mathbf{k}_3 \neq \mathbf{k}_{1,2}$,  then it is easy to check that 
\begin{gather}
   \psi_{\mathbf{k}_3}\times \psi_\mathbf{k} = \psi_{\mathbf{k}_1} \left(\psi_{\mathbf{k}_2} \cdot \psi_{\mathbf{k}_3}\right) -  \psi_{\mathbf{k}_2}  \left(\psi_{\mathbf{k}_1} \cdot \psi_{\mathbf{k}_3}\right)  = 0\,,
\end{gather}
where we have used that $\psi_{\mathbf{k}_{1,2}} \cdot \psi_{\mathbf{k}_3} = 0$. But it means that $\psi_{\mathbf{k}_3} = \mathcal{C}(\mathbf{r}) \psi_\mathbf{k}$ for some function $\mathcal{C}(\mathbf{r})$ (two spinors are collinear). Since $\mathbf{k}_3$ can be  arbitrary we conclude that each wave function in the rank $2$ flat band is linearly dependent to each other.  That is in contradiction with our initial assumption that the flat band has rank $2$. Therefore we  conclude that for $\textbf{d}=0$ the flat bands can not realize \textit{Scenario 2} and thus have the Chern number $C=2$. 

\begin{figure}[t!]
    \centering
    \includegraphics[width=0.42\textwidth]{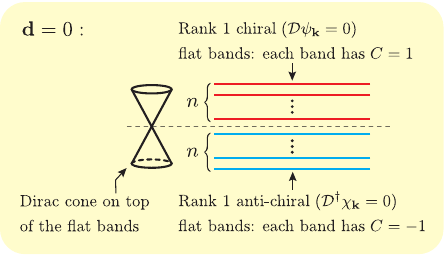}
    \caption{A structure of  the zero energy flat bands in  the case of  the displacement $\mathbf{d} = 0$.}
    \label{fig:strfld0}
\end{figure}

\textit{Dirac cone on top of the flat bands.}
Now we  prove that if we have rank $1$ flat bands then we must have a Dirac cone on top of them. Namely we prove that we have an additional zero energy mode of the chiral  (and thus also zero energy mode of the anti-chiral) operator  at an isolated point of the mBZ.  We consider a case of a single chiral flat band, when the corresponding wave function  $\psi_\mathbf{k}(\mathbf{r})$ has only one zero. We can pick such $\mathbf{k}_0$ for which $\psi_{\textbf{k}_{0}}(0)=0$. Using discussion below (\ref{vhol}) one can notice that $\psi_{\mathbf{k}_0}(\textbf{r}) \cdot  \psi_{\mathbf{k}_0}(-\mathbf{r}) = 0$. Due to this fact the following equation
\begin{gather}
    \psi_{\mathbf{k}_0}(\textbf{r}) = \hat{\phi}_{\mathbf{k}_{*}-2\mathbf{k}_0 }(-\textbf{r}) \times \psi_{\mathbf{k}_0}(-\textbf{r})
\end{gather}
has a finite solution $\hat{\phi}_{\mathbf{k}_{*}-2\mathbf{k}_0}$ at the point $\mathbf{k}_{*} - 2\mathbf{k}_{0}$ of the mBZ.
Then if we apply the operator $\mathcal{D}(\mathbf{r})$ to both sides of this equation we get
\begin{gather}
    0 = \left(\mathcal{D}(-\mathbf{r})\hat{\phi}_{\mathbf{k}_{*}-2\mathbf{k}_0 }(-\mathbf{r})\right) \times \psi_{\mathbf{k}_0}(-\mathbf{r})\,,
\end{gather}
which leads to
\begin{gather}
\mathcal{D}(\mathbf{r})\hat{\phi}_{\mathbf{k}_{*}-2\mathbf{k}_0 }(\mathbf{r}) =\mathcal{C}(\mathbf{r})\psi_{\mathbf{k}_{*}-2\mathbf{k}_0 }(\mathbf{r})\,.
\end{gather}
Since $\phi_{\mathbf{k}_{*}-2\mathbf{k}_0}$ has a zero at some point we can always find a periodic function $g(\mathbf{r})$ \cite{PhysRevB.103.155150} such that
\begin{gather}
    \left(\mathcal{C}(\mathbf{r}) + \bar{\partial} g(\mathbf{r})\right) \psi_{\mathbf{k}_{*}-2\mathbf{k}_0 }(\mathbf{r}) = 0
\end{gather}
and therefore we can construct a solution $\phi_{\mathbf{k}_{*}-2\mathbf{k}_0 }(\mathbf{r})  = \hat{\phi}_{\mathbf{k}_{*}-2\mathbf{k}_0 }(\mathbf{r}) +g(\mathbf{r}) \psi_{\mathbf{k}_{*}-2\mathbf{k}_0 }(\mathbf{r}) $ that belongs to a specific point of the mBZ and satisfies the equation
\begin{gather}
    \mathcal{D}(\mathbf{r})\phi_{\mathbf{k}_{*}-2\mathbf{k}_0 }(\mathbf{r})  = 0\,.
\end{gather}
This concludes the classification of the flat bands structures for $\mathbf{d}=0$. In summary, we can have only chiral and anti-chiral flat bands of rank $1$ with the Chern numbers $C=1$ and $C=-1$ with a single Dirac cone on top of them. This is schematically depicted in Fig. \ref{fig:strfld0}. 
\vspace{0.5cm}

\begin{figure}[t!]
    \centering
    \includegraphics[width=0.48\textwidth]{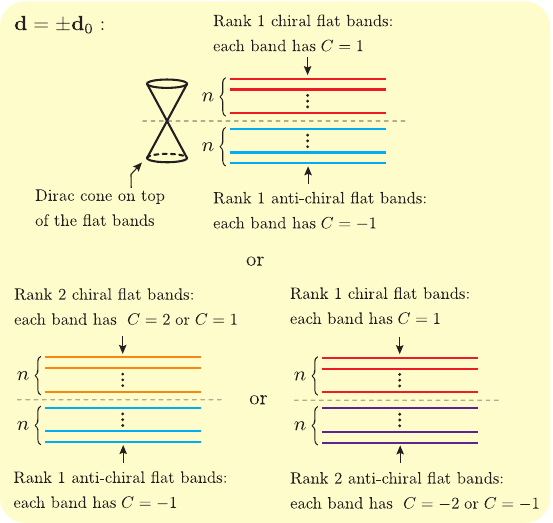}
    \caption{Possible structures of  the zero energy flat bands in the case of  the displacement $\mathbf{d} = \pm \textbf{d}_{0}$.}
    \label{fig:strfld1}
\end{figure}
\vspace{0.5cm}

\subsection{Displacement $\textbf{d}=\pm\textbf{d}_{0}$}
\textit{Possible topologies of the flat bands.} In the case of non-zero displacement $\textbf{d}=\pm \textbf{d}_0$ we can construct an anti-chiral zero mode  by taking cross product of two chiral zero modes: 
\begin{gather}
    \chi_{\mathbf{k}_{*} - \mathbf{k}_1 - \mathbf{k}_2}(\mathbf{r}) = Q\left(\bar{\psi}_{\mathbf{k}_1}(\mathbf{r}) \times\bar{\psi}_{\mathbf{k}_2}(\mathbf{r})\right)\,.\label{eq:ABAprodformula}
\end{gather}
Even though we can not construct an anti-chiral mode from a chiral one using complex conjugation and inversion as in (\ref{eq:maps}), the number of anti-chiral flat bands must be equal to the number of the chiral ones, because of the particle-hole symmetry. 

Also similarly  to the $\textbf{d}=0$ case one can show that the following function  
\begin{gather}
    v(\mathbf{r}) =   \bar{\chi}_{\mathbf{k}_1}(-\mathbf{r})Q \psi_{\mathbf{k}_2}(\mathbf{r}) \label{vhol2}
\end{gather}
satisfies $\bar{\partial} v(\mathbf{r})  = 0$ and thus  is equal to zero if $\mathbf{k}_1 \neq \mathbf{k}_2$. 


Unlike the case  $\textbf{d}=0$, when $\textbf{d}=\pm \textbf{d}_{0}$ there are no obstacles preventing the existence of rank $2$ flat bands. Let us assume that we have a rank $2$ chiral flat band, thus  we have two  linearly independent (besides $n$ points) wave spinors $\psi_{\mathbf{k}_1}$ and $\psi_{\mathbf{k}_2}$. Using \eqref{eq:ABAprodformula} we construct an anti-chiral zero mode $\chi_\mathbf{k}$. The number of zeros of $\chi_{\mathbf{k}}$ is equal to the number of points where $\psi_{\mathbf{k}_1}$ and $\psi_{\mathbf{k}_2}$ are linearly dependent. If we assume that there is some other anti-chiral zero mode $\chi_{\mathbf{k}_3}$ with $\mathbf{k}_3 \neq \mathbf{k}_1,\mathbf{k}_2$  we have
\begin{gather}
    \chi_{\mathbf{k}}(\textbf{r}) \times \chi_{\mathbf{k}_3}(\textbf{r}) = Q\left(Q\left(\psi_{\mathbf{k}_1}\times \psi_{\mathbf{k}_1} \right) \times \bar{\chi}_{\mathbf{k}_3}\right) = 0\,,
\end{gather}
where we used (\ref{vhol2}). 
Therefore the anti-chiral flat band must have  rank $1$. From that it follows that we can not have at the same time rank $2$ chiral and anti-chiral flat bands. Also one can prove that the Dirac cone on top of the flat bands is prohibited in this case.

Finally, if we assume that we have a rank $1$ chiral flat band $\psi_{\textbf{k}}$ and one Dirac cone on top of it with its zero mode $\phi_{\textbf{k}_{0}}$. Then using \eqref{eq:ABAprodformula} for $\phi_{\textbf{k}_{0}}$ and $\psi_{\textbf{k}}$ we can construct  a rank $1$ anti-chiral flat band. And vice versa, using the same argument as in the previous subsection we can prove that if we have rank $1$ chiral and anti-chiral flat bands, we must have a single Dirac cone on top of them. We show schematically  in Fig. \ref{fig:strfld1} all the possible cases for $\textbf{d}=\pm \textbf{d}_{0}$.

\section*{Acknowledgments}  

We are grateful to Simon Becker, Tristan Humbert, Igor R. Klebanov, Jens Wittsten, Mengxuan Yang  for useful discussions. 
We would like to thank Igor R. Klebanov for valuable comments on the draft. F.K.P. is currently a Simons Junior Fellow at New York University and supported by a grant 855325FP from the Simons Foundation.

\bibliographystyle{ieeetr} 
\bibliography{Trilayer} 


\end{document}